\documentclass{aa}

\usepackage{graphicx}
\usepackage{txfonts}
\usepackage{color}

\begin{document}

\title{Multiple stellar populations in Schwarzschild modeling\\ and the application to the Fornax dwarf}

\author{Klaudia Kowalczyk and Ewa L. {\L}okas
}

\institute{Nicolaus Copernicus Astronomical Center, Polish Academy of Sciences,
Bartycka 18, 00-716 Warsaw, Poland\\
\email{klaudia.kowalczyk@gmail.com, lokas@camk.edu.pl}}


\titlerunning{Multiple stellar populations in Schwarzschild modeling}
\authorrunning{K. Kowalczyk \& E. L. {\L}okas}

\abstract{
Dwarf spheroidal (dSph) galaxies are believed to be strongly dark matter dominated and thus are
considered perfect objects to study dark matter distribution and test theories of structure formation. They possess
resolved, multiple stellar populations that offer new possibilities for modeling. A promising tool for the dynamical
modeling of these objects is the Schwarzschild orbit superposition method. In this work we extend our previous
implementation of the scheme to include more than one population of stars and a more general form of the mass-to-light
ratio function. We tested the improved approach on a nearly spherical, gas-free galaxy formed in the cosmological
context from the Illustris simulation. We modeled the binned velocity moments for stars split into two populations by
metallicity and demonstrate that in spite of larger sampling errors the increased number of constraints leads to
significantly tighter confidence regions on the recovered density and velocity anisotropy profiles. We then applied the
method to the Fornax dSph galaxy with stars similarly divided into two populations. In comparison with our earlier
work, we find the anisotropy parameter to be slightly increasing, rather than decreasing, with radius and more strongly
constrained. We are also able to infer anisotropy for each stellar population separately and find them to be
significantly different.
 }


\maketitle

\section{Introduction}
\label{sec:intro}

Dwarf spheroidal (dSph) galaxies of the Local Group \citep{mateo_1998, tolstoy_2009} are considered to be a perfect
tool to test our current theories of structure formation involving dark matter in the context of near-field cosmology.
The objects are believed to be strongly dark matter dominated with mass-to-light ratios even on the order of a few
hundred solar units. Due to their proximity they are also the only extragalactic systems where individual stars can be
resolved and their velocities measured offering the possibility to create interesting dynamical modeling techniques.

The first estimates of dark matter content in dSph galaxies were based on a single measurement of the line-of-sight
velocity dispersion of the stars and the application of the virial theorem. As the samples of the stars with kinematic
measurements grew, it became possible to estimate the profile of the velocity dispersion and model it using the Jeans
equation \citep{GD}. Since the stars in the galaxy can move on a variety of orbits, from circular to radial, the
degeneracy between the anisotropy of the orbits and the mass distribution is inherent in this type of modeling. The
reason for this lies in the fact that different combinations of these quantities can reproduce the velocity dispersion
profile equally well.

A way to overcome this issue, at least partially, is to resort to higher order line-of-sight velocity moments, such as
the kurtosis, and use the corresponding Jeans equations. Since the kurtosis is more sensitive to the velocity
anisotropy than to the mass distribution, useful constraints can be obtained on both. Still, the method requires large
kinematic samples to estimate the velocity moments reliably and some assumption on the functional form of the
anisotropy \citep{lokas_2002, lokas_2005}.

The Schwarzschild modeling technique \citep{schwarzschild_1979} offers a different approach to estimate the properties
of dSph galaxies without prior assumptions on the type of orbits. It relies on building a galaxy model out of a set of
best-fitting orbits probed in the range of energy and angular momenta. In this method, the anisotropy of the stellar
orbits comes out as a result of the modeling in the same way as the density profile. Although it has been originally
developed for large elliptical galaxies \citep{marel_1998, valluri_2004, gebhardt_2003}, it has recently been adopted
for use on discrete data characteristic of dSph galaxies and applied to a number of dwarfs, including Carina, Draco,
Fornax, Sculptor, and Sextans \citep{jardel_2012, jardel_2013, breddels_2013, breddels_2013a, kowalczyk_2019}.

Many dSph galaxies show signs of the presence of multiple stellar populations resulting from a few star formation
episodes \citep{bellazzini_2001, delpino_2015, fabrizio_2016, pace_2020}. This observation offers a way to improve the
modeling methods since, assuming dynamical equilibrium, all populations are supposed to be influenced by the same
underlying gravitational potential of the galaxy, but they have different distributions so more constraints can be
imposed during the modeling. This approach was first used by \citet{battaglia_2008} to model the mass distribution
in the Sculptor dSph galaxy. A few attempts have also been made to constrain the inner slope of the dark matter profile
in dSph galaxies using this technique \citep{walker_2011, amorisco_2012, hayashi_2018} in order to resolve the
so-called cusp-core problem. It has been shown to be difficult, however, due to the nonsphericity of the dwarfs that
introduces biases in such measurements \citep{kowalczyk_2013, genina_2018}.

In our recent papers \citep{kowalczyk_2017, kowalczyk_2018, kowalczyk_2019} we developed the Schwarzschild technique
in the form applicable to binned velocity moments of a single tracer and verified its ability to reproduce the mass
distribution and velocity anisotropy of simulated galaxies. We have also studied biases resulting from the
nonsphericity of the modeled objects. Later, we applied the method to model the kinematics of the Fornax dSph galaxy
estimating its mass and anisotropy profiles with unprecedented precision.

In this paper we extend our Schwarzschild modeling technique to include multiple stellar populations with the aim to
constrain the properties of dSph galaxies even more strongly. We test our approach on a realistic simulated galaxy
formed in the cosmological context, originating from the Illustris project \citep{vogelsberger_2014a}. Although no
precise analogues of dSph galaxies are available in this simulation because of the resolution, we use a more massive
galaxy but with properties otherwise similar to dSphs. The reliability of the modeling does not depend on the particular
value of the mass so we believe these tests to be viable. We do not attempt to constrain the inner dark matter density
profile (which is poorly resolved anyway) but try to put tighter limits on the estimates of the mass and anisotropy
profiles. Finally, we apply the improved method to the available kinematic data for the distinct stellar populations of
the Fornax dSph.

This paper is organized as follows. In Section~\ref{sec:data} we present the data for the simulated galaxy as well as
their splitting into stellar populations and mock observations along the main axes. Section~\ref{sec:model} contains an
overview of our modeling method, the application of the method to all stars and to two populations, and a comparison of
the results obtained with these two approaches. The results of the application of the method to the Fornax dSph galaxy
are presented in Section~\ref{sec:fornax}. We discuss our findings and summarize the paper in
Section~\ref{sec:summary}.

\begin{table}
\caption{Properties of the Illustris galaxy used to create mock data.}
\label{tab:subhalo}
\centering
\begin{tabular}{l c}
\hline\hline
Property & Value  \\ \hline
Subhalo ID & 16960 \\
Number of stellar particles ($N_{\star}$) & 70446 \\
Number of dark matter particles ($N_{\rm DM}$) & 78448 \\
Stellar mass ($M_{\star}$) & $5.74\times 10^{10}\,\rm{M}_{\odot}$ \\
Dark matter mass ($M_{\rm DM}$) & $4.91\times 10^{11}\,\rm{M}_{\odot}$ \\
Mean mass of stellar particles & $815808\,\rm{M}_{\odot}$ \\
Stellar half-mass radius & $9.99\,\rm{kpc}$ \\
Stellar half-number radius ($r_{1/2}$) & 9.6\,\rm{kpc}\\
Axis ratio $c/a$ within $r_{1/2}$ & 0.907\\
Axis ratio $b/a$ within $r_{1/2}$ & 0.949\\
Triaxiality & 0.56 \\
\hline
\end{tabular}
\end{table}

\begin{figure}[ht!]
\includegraphics[width=\columnwidth]{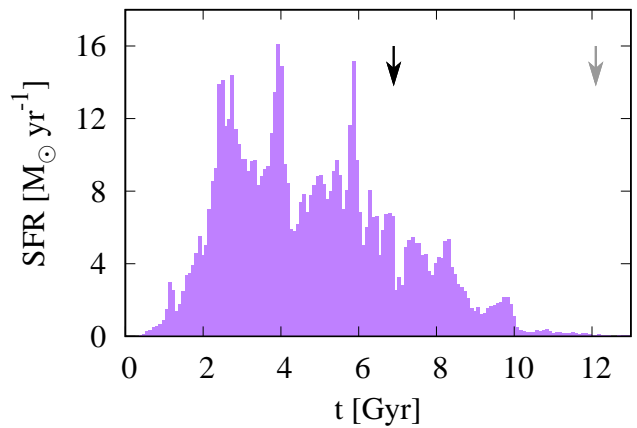}
\caption{
Star formation rate as a function of the age of the Universe in the simulated galaxy from the Illustris
project used to create mock data. The black and gray vertical arrows indicate the last mergers which the
galaxy underwent, wet and dry, respectively.}
\label{fig:sfr}
\end{figure}

\begin{figure}[ht!]
\hspace{0.3cm}
\includegraphics[width=\columnwidth]{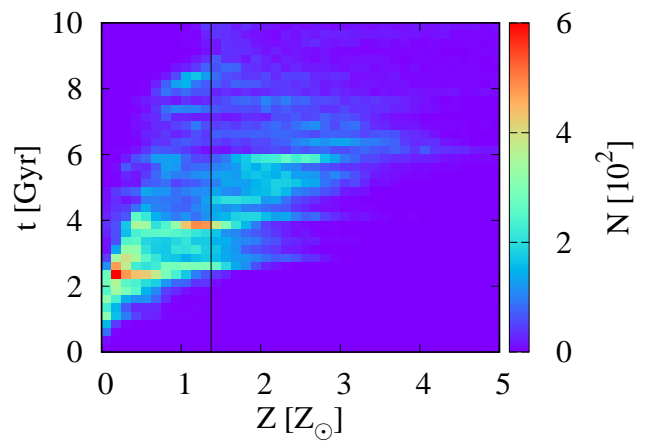}
\caption{Number of stars as a function of their metallicity and
time of formation (the age of the Universe) in the simulated galaxy. The vertical line indicates the applied
split into stellar populations.}
\label{fig:metal}
\end{figure}

\begin{figure*}[ht!]
\includegraphics[width=\textwidth]{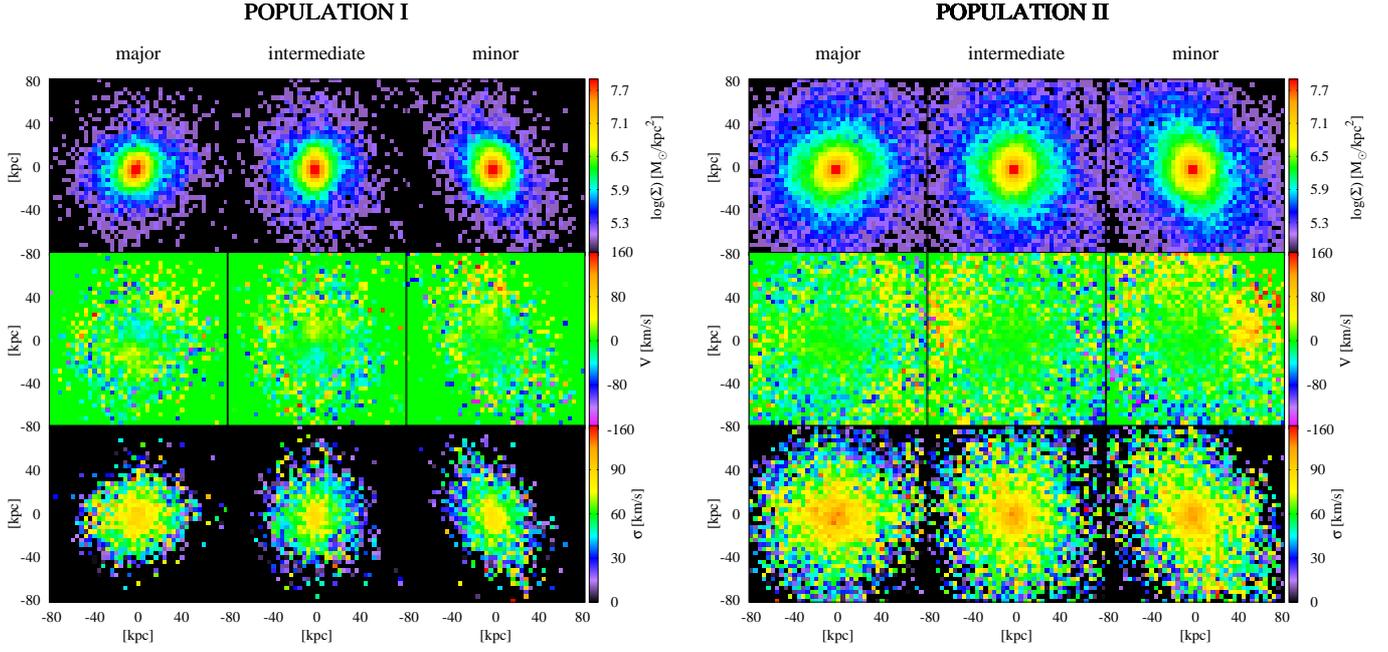}
\caption{Maps of the projected stellar density, mean stellar velocity, and stellar velocity dispersion (in rows)
for two stellar populations: the metal-rich population I (left-hand side panels) and the metal-poor population II
(right-hand side), and observations along the principal axes determined for all stars (in columns, along the
major, the intermediate, and the minor axis, respectively).}
\label{fig:maps}
\end{figure*}

\begin{figure*}[ht!]
\begin{center}
\includegraphics[width=\textwidth]{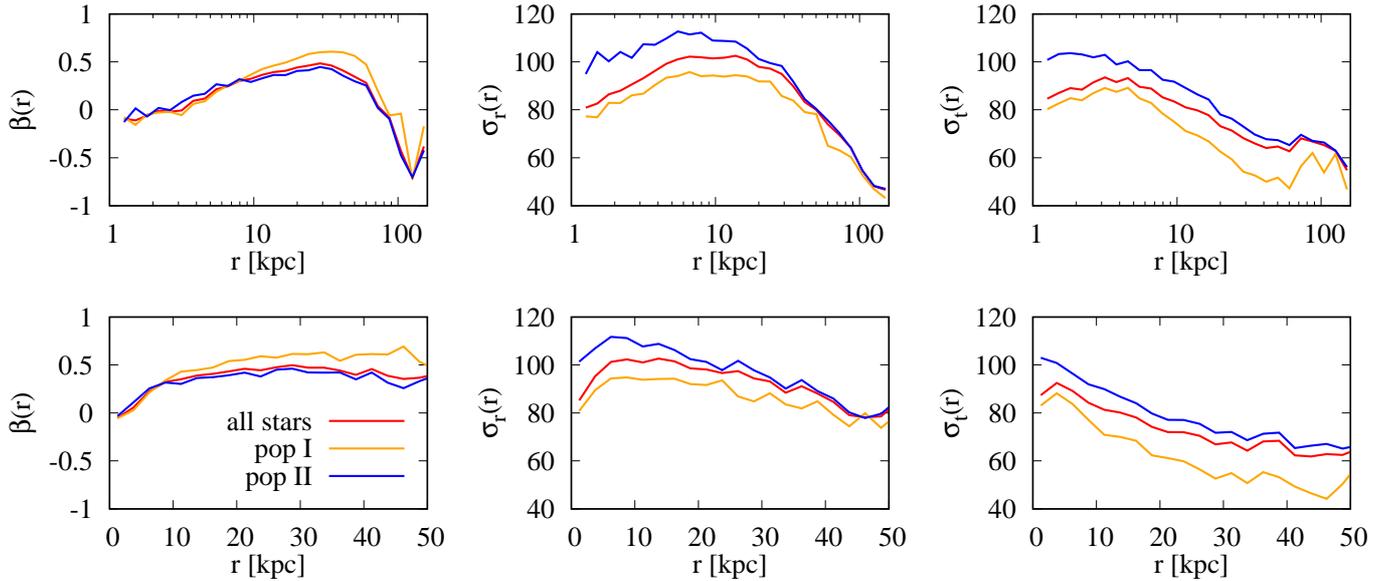}
\caption{Profiles of the velocity anisotropy parameter, radial velocity dispersion, and tangential velocity
dispersion (in consecutive columns) calculated from all stars (in red), including only population I (in orange), and
only population II (in blue). The upper row shows the profiles using the logarithmic distance scale and reaching the
outskirts of the galaxy whereas the bottom row presents in the linear scale only the radial range used in the
modeling.}
\label{fig:beta}
\end{center}
\end{figure*}

\begin{figure*}[ht!]
\begin{center}
\vspace{0.5cm}
\includegraphics[width=0.9\textwidth]{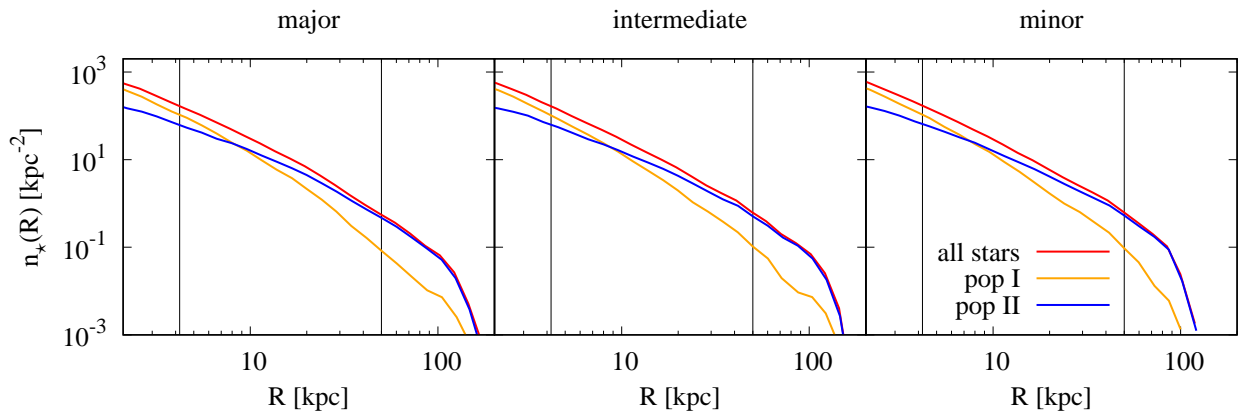}
\caption{Surface number density profiles of the stellar data samples for the simulated galaxy observed along different
lines of sight (from the left to the right). Different lines show profiles for all available stars (in red),
the metal-rich population I (in orange), and the metal-poor population II (in blue). Thin vertical lines indicate $r_0$
(see text) and the outer boundary of the spectroscopic data.}
\label{fig:profile}
\end{center}
\end{figure*}

\begin{figure*}[ht!]
\begin{center}
\vspace{-1.0cm}
\includegraphics[width=0.85\textwidth]{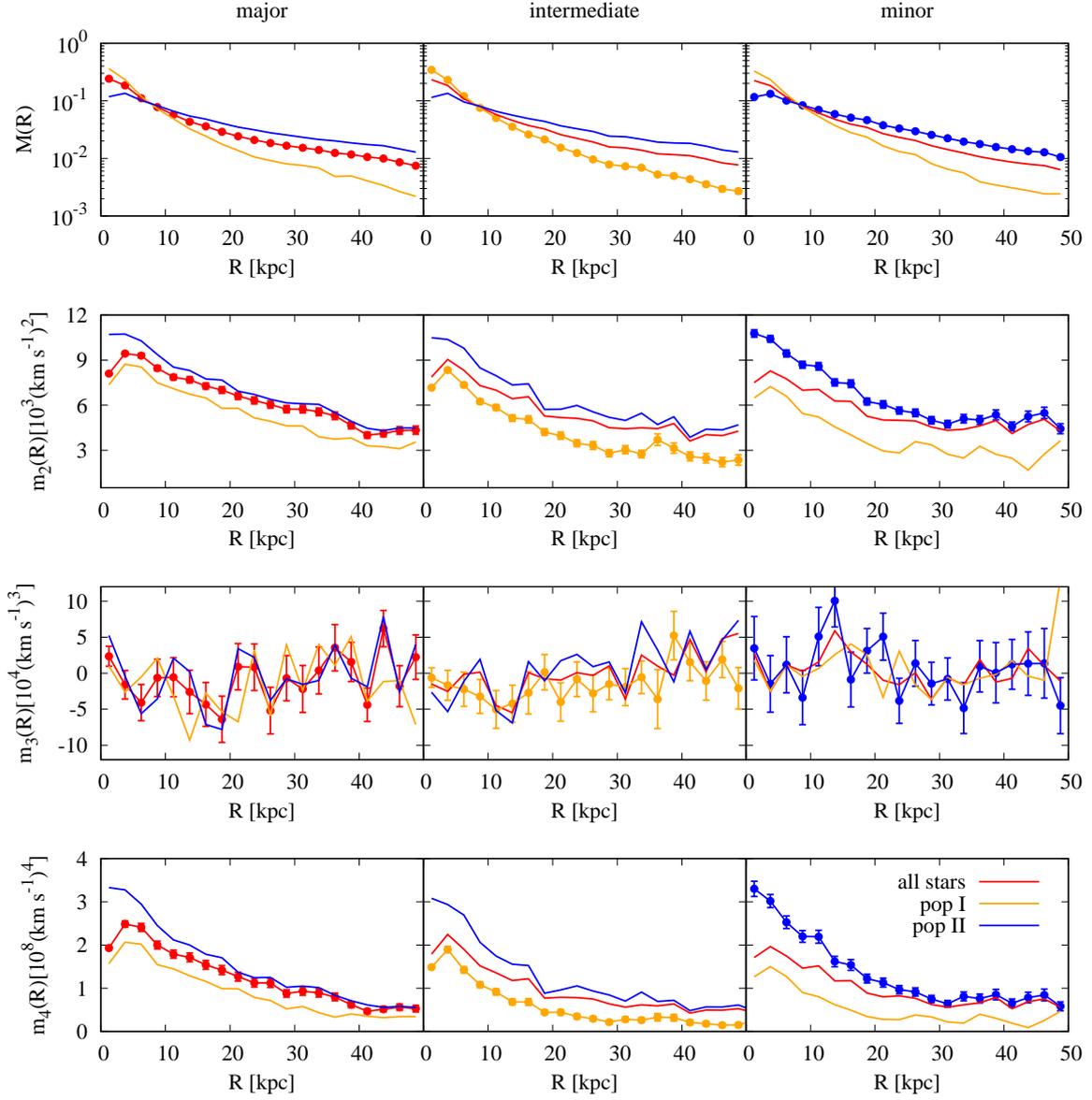}
\caption{Observables used in our Schwarzschild modeling scheme of the simulated galaxy. \textit{In rows:} the
fraction of the total number of stars, 2nd, 3rd, and 4th velocity moment. \textit{In columns:} mock data from the
simulated galaxy along the major, intermediate, and minor axis. In red we present the values obtained for all stars
whereas in orange and blue those for populations I and II, respectively. For clarity of the figure, in each panel we
indicate only the error bars for one of the data sets.}
\label{fig:obs}
\end{center}
\end{figure*}

\begin{figure*}[ht!]
\centering
\includegraphics[trim=0 0 0 270, clip, width=0.9\textwidth]{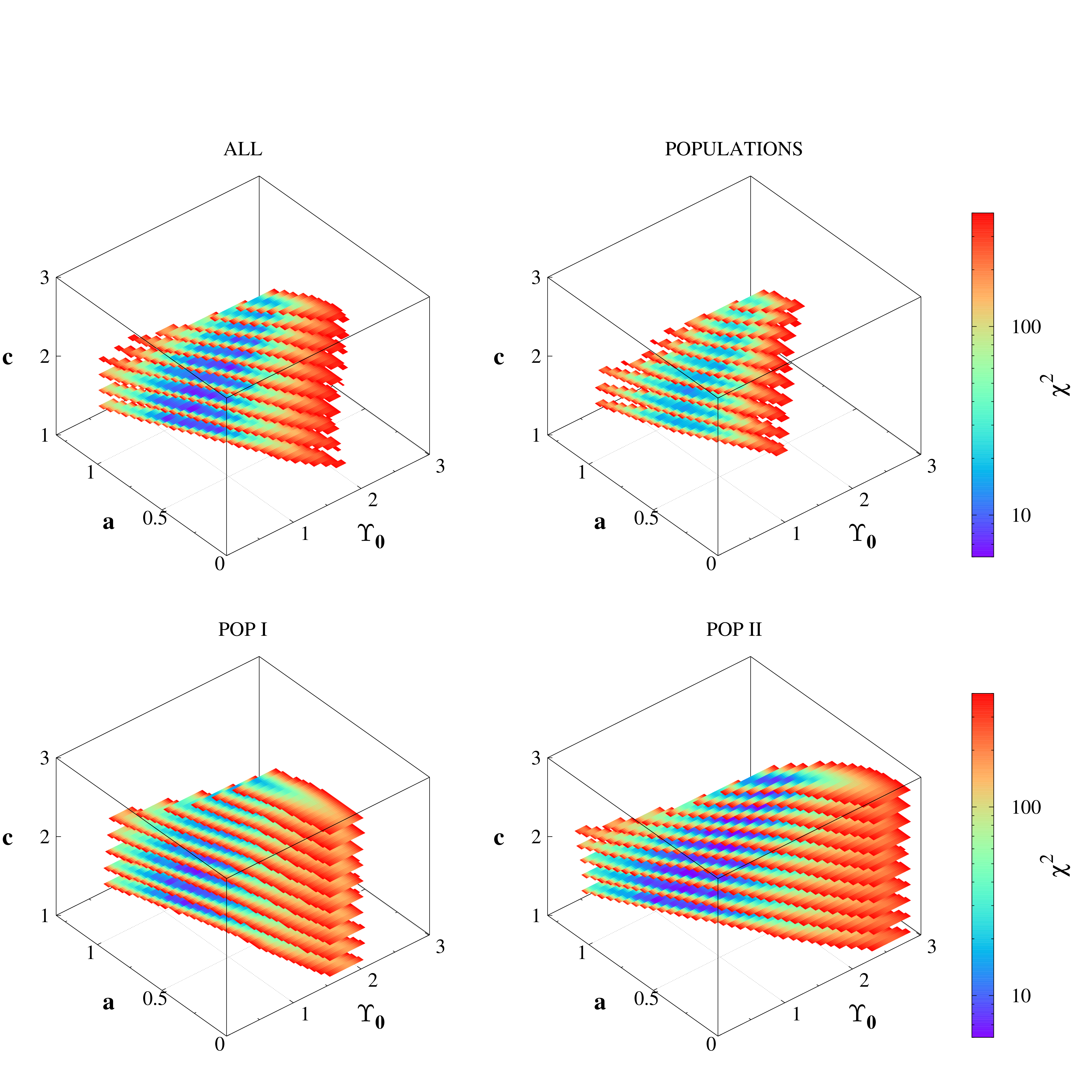}
\caption{Absolute values of $\chi^2$ obtained from the fits of three data sets: all stars (top left panel), population
I (bottom left), and population II (bottom right) for the observations along the major axis of the
simulated galaxy. The results for the modeling of two populations (top right) were obtained as an algebraic sum of
values for populations I and II. To avoid large numbers in the figure, $\Upsilon_0$ was divided by the mean mass of a
stellar particle.}
\label{fig:chi}
\end{figure*}

\section{Mock data}
\label{sec:data}

\subsection{Selection of the simulated galaxy}

In order to test our modeling method on realistic simulated data, we decided to use a galaxy from the Illustris
project \citep{vogelsberger_2014a, vogelsberger_2014b, genel_2014, nelson_2015}, namely the Illustris-1 cosmological
simulation. This simulation follows the formation and evolution of galaxies from the early Universe to the present by
solving gravity and hydrodynamics, as well as modeling of star formation, galactic winds, magnetic fields, and
the feedback from black holes. Although dwarf galaxies that are of our interest here are not resolved in the suite, this
can be easily overcome with the appropriate choice of the object and the treatment of data.

As the key properties of dSph galaxy equivalents we identified: the lack of gas, the lack of a black hole, a low spin,
the stellar mass much smaller than the dark matter mass and a nearly spherical shape. The last condition was adopted in
an attempt to avoid any strong bias introduced by the spherical modeling of a nonspherical object. Moreover, we
required the galaxy to possess a significant number of both stellar and dark matter particles (over $10^5$), and a well
resolved center. Due to the large softening scale for dark matter particles in the simulation
($\epsilon_{\rm DM}=1.42\,$kpc), we looked for an object in which even the more concentrated stellar population (see
Section~\ref{sec:pops}) extended over 43\,kpc so that the region affected by the numerical artifacts was enclosed within
2-3 innermost data bins (we used 20 linearly spaced spatial bins, see Section~\ref{sec:method}).

Out of 27345 galaxies listed in the catalog of stellar circularities, angular momenta, and axis ratios published by
the Illustris team \citep{genel_2015} containing subhalos with the stellar mass larger than $10^9$\,M$_{\odot}$, only
a few met our restrictive requirements. We decided to use a galaxy labeled as subhalo 16960. All the relevant
properties of the galaxy are given in Table\,\ref{tab:subhalo}, including numbers of particles and total masses for both
components, and details on the shape of the stellar component: the axis ratios minor to major (shortest to longest)
$c/a$, intermediate to major $b/a$, and the triaxiality parameter $T=(a^2-b^2)/(a^2-c^2)$. We distinguish between the
half-mass radius provided in the Illustris database and the half-number radius $r_{1/2}$, which we use for further
calculations in this paper. The difference between the two comes from a small gradient in the stellar mass-to-light
ratio with the distance from the galactic center. Since in our approach we treat stars as equal-mass particles and
refer to number densities (multiplied by the mean mass of a stellar particle when needed), the application of the
half-number radius is more self-consistent.

\begin{figure*}[ht!]
\includegraphics[width=\columnwidth]{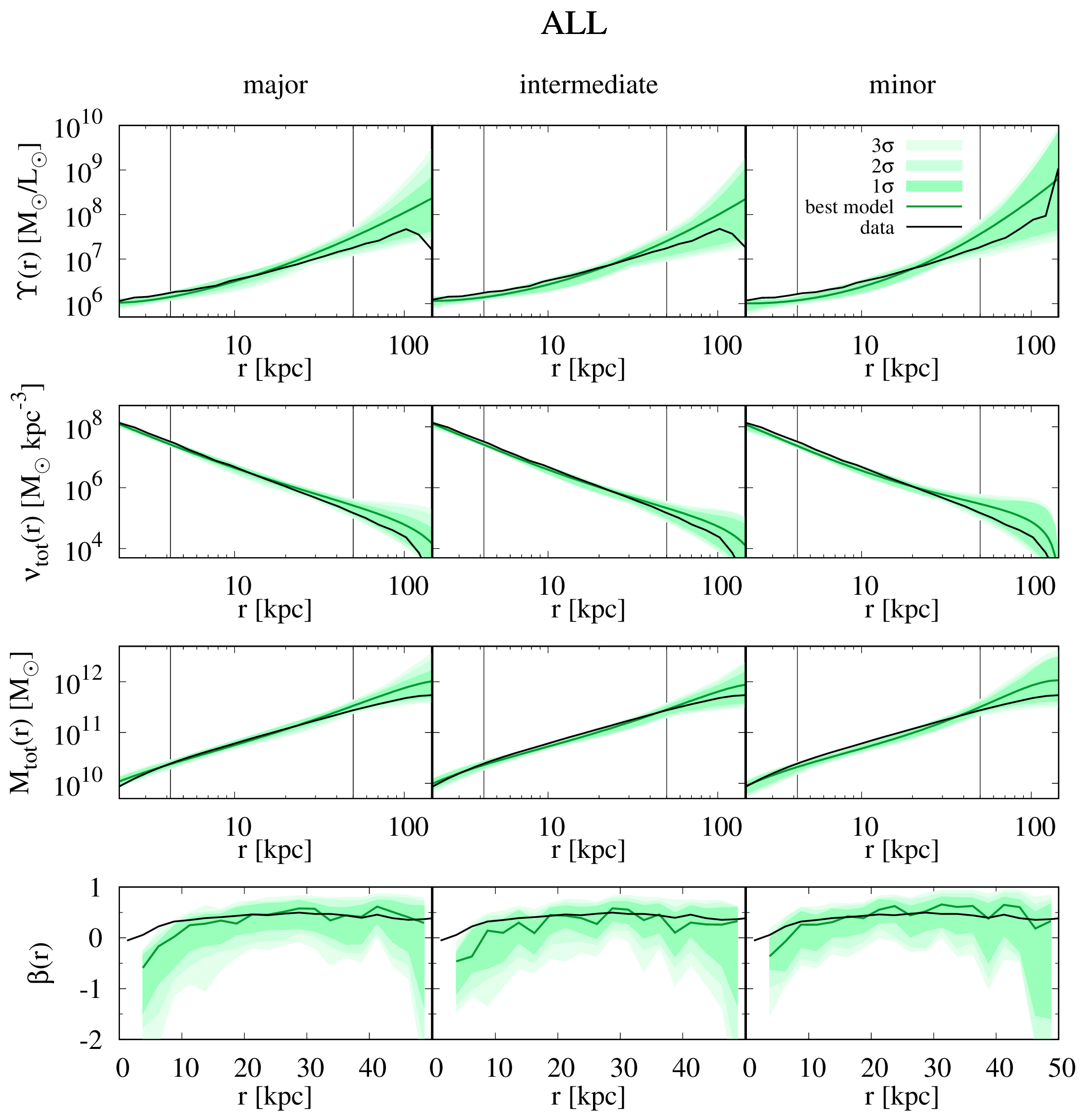}
\hspace{0.2cm}
\includegraphics[width=\columnwidth]{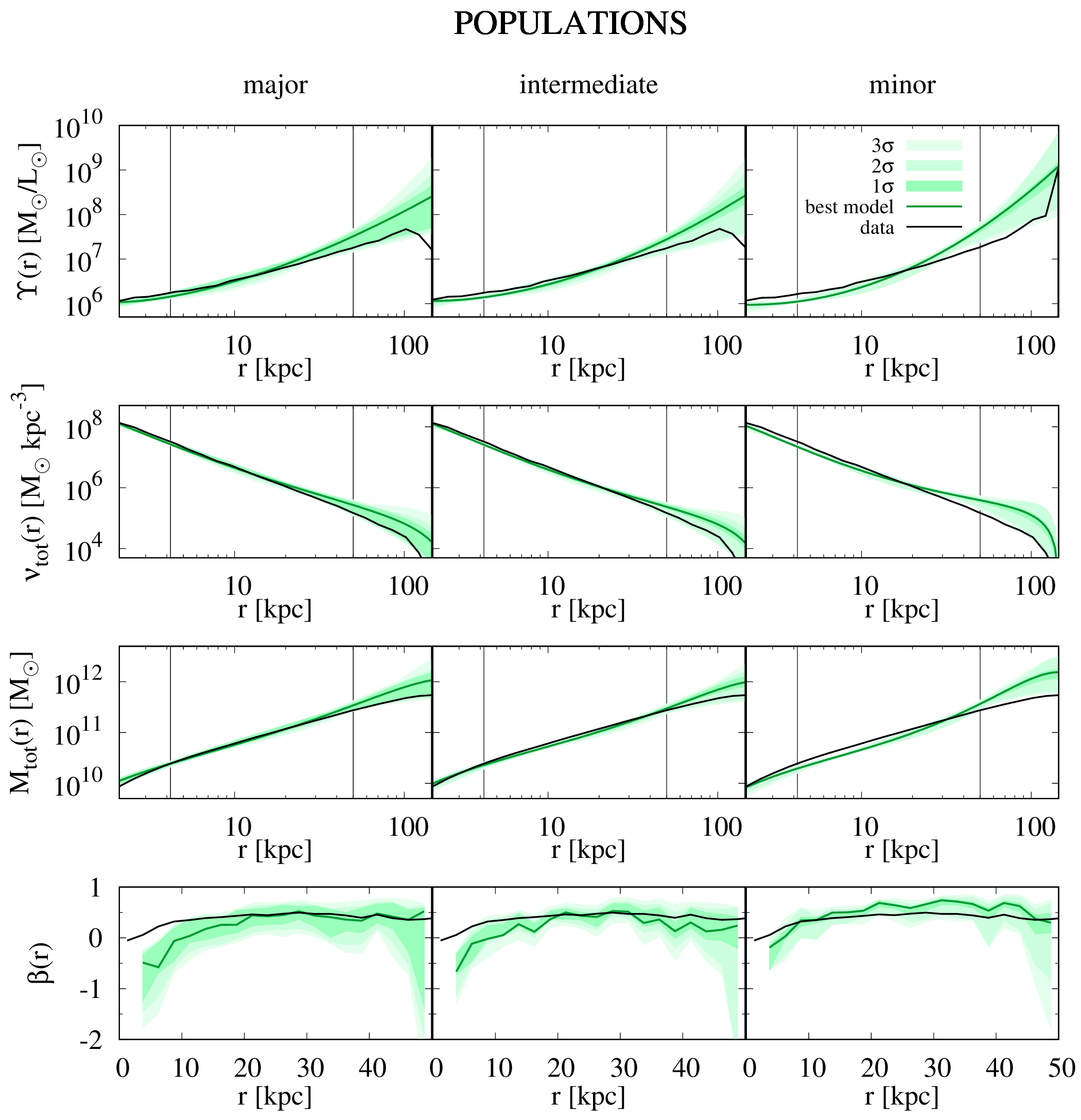}
\caption{Left-hand side: results of Schwarzschild modeling of three mock data sets obtained by observing the
simulated galaxy along the principal axes. \textit{In rows:} derived mass-to-light ratio, total density, total mass, and
anisotropy parameter. \textit{In columns:} observations along the major, intermediate, and minor axis,
respectively. Green lines indicate values for the best-fit models whereas the colored areas of decreasing intensity
show the 1, 2, and 3\,$\sigma$ confidence levels. The true values are presented as black lines. Thin vertical lines mark
the values of $r_0$ and the outer range of the data sets, from left to right. Right-hand side: same as
left but for the fit of two stellar populations.}
\label{fig:ups}
\end{figure*}

\subsection{Splitting the stars into populations}
\label{sec:pops}

Our chosen galaxy shows a complex formation history undergoing multiple mergers which result in extended star
formation with a few star formation bursts. The last wet merger, that is a merger with an object containing gas,
happens at 6.9\,Gyr from the beginning of the simulation, whereas the last dry merger (no gas transfer)
at 12.1\,Gyr, giving the galaxy enough time to regain dynamical equilibrium. We present the star
formation rate (SFR) as a function of time (the age of the Universe) in Fig.\,\ref{fig:sfr}, where these last
mergers are indicated with black and gray vertical arrows. In Fig.\,\ref{fig:metal} we show the distribution of stars
as a function of their metallicity (in solar units) and the time of formation. In order to divide the stellar sample
into two populations we cut it in half based on the metallicity index of each stellar particle. This split is indicated
in Fig.\,\ref{fig:metal} with the vertical line. With satisfying accuracy it separates the stars born before and after
4 Gyr since the start of the simulation, which corresponds to the formation time before and after the end of the second
major star burst, as shown in Fig.\,\ref{fig:sfr}. We refer to the metal-rich stars as population I and to the
metal-poor as population II, following the commonly used nomenclature in astronomy.

In Fig.\,\ref{fig:maps} we present maps of the projected stellar mass density, line-of-sight velocity, and line-of-sight
velocity dispersion for both populations obtained by projecting the galaxy along its principal axes. The orientation
was determined from the inertia tensor calculated from all stars within the half-number radius $r_{1/2}$ and therefore
is the same in both panels. The two populations differ significantly in the spatial distribution and kinematics with the
metal-rich (considered to be younger) population I being more concentrated but having lower central velocity
dispersion. Both populations show a weak rotation signal at large distances from the center.

The velocity anisotropy parameter $\beta (r) = 1 - (\sigma_\theta^2 + \sigma_\phi^2)/(2 \sigma_r^2)$, where $\sigma_i$
are velocity dispersions in spherical coordinates \citep{GD}, describes the orbital structure of galaxies. It is one of
the most important dynamical properties of bound systems which cannot be inferred directly from observations and has to
be recovered by dynamical modeling. The profiles of the anisotropy parameter $\beta$ as well as the radial
$\sigma_r$ and tangential $\sigma_t = [(\sigma_\theta^2 + \sigma_\phi^2)/2]^{1/2}$ velocity dispersions for our
simulated galaxy are presented in the consecutive columns of Fig.\,\ref{fig:beta}. Throughout the paper we use red,
orange, and blue colors to indicate values calculated or recovered for all stars, population I, and population II,
respectively. The two rows of the figure show the behavior of the parameters at different scales. The top row plots the
profiles with the distance from the center of the galaxy in the logarithmic scale and shows the drop of anisotropy at
the outer edges of the object. The bottom row uses the linear distance scale and focuses on the main body of the
galaxy.

Figure~\ref{fig:profile} shows the surface number density profiles of the stars as measured
in different directions. We can see that while the different subsamples have quite distinguishable profiles, the
difference between the lines of sight is small because the galaxy is close to spherical.

\begin{figure*}[ht!]
\centering
\includegraphics[width=0.9\textwidth]{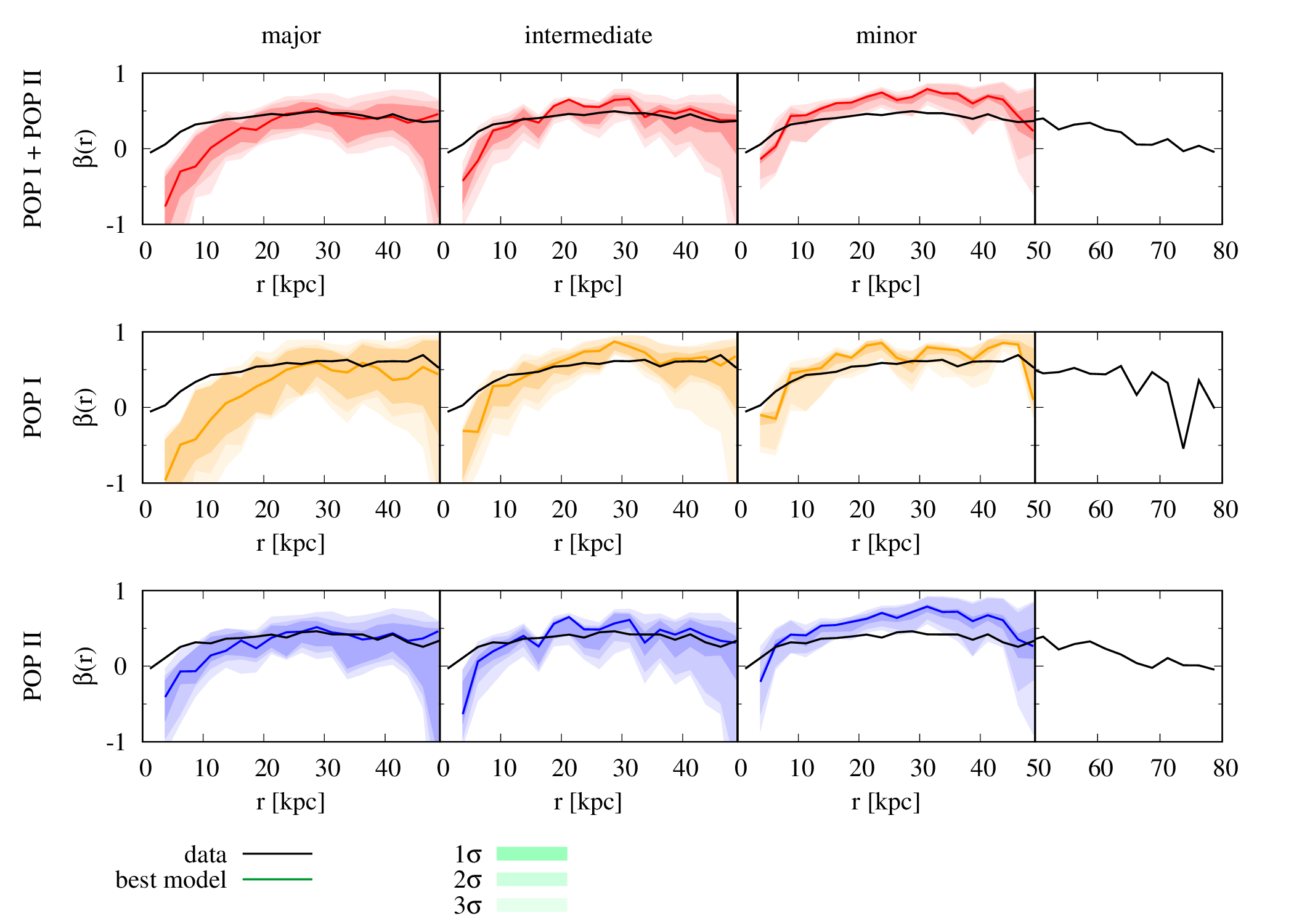}
\caption{Profiles of the anisotropy parameter obtained with the Schwarzschild modeling of two stellar
populations of the simulated galaxy. \textit{In rows:} results for all stars (calculated as the superposition of two
populations), population I, and population II. Colors follow the convention used in previous figures. \textit{In
columns:} observations along the major, intermediate, and minor axis. The last narrower column shows the data (black
lines) outside the modeled radial range. Color lines indicate values for the best-fit models whereas the colored areas
of decreasing intensity show the 1, 2, and 3\,$\sigma$ confidence regions.}
\label{fig:beta_fit}
\end{figure*}

\subsection{Observables}
\label{sec:obs}

We generated nine sets of mock data by observing all stars and each population separately along the principal axes
determined from all stars. For the observables to be used
in the modeling we divided the stars into 20 bins spaced linearly in distance from the center of the galaxy up to
50\,kpc, measuring the fraction of the total number of stars and the 2nd, 3rd, and 4th proper moments of the
line-of-sight velocity defined in Eq.\,8 and 9 of \citet{kowalczyk_2018}. The profiles of these quantities are shown in
consecutive rows in Fig.\,\ref{fig:obs}. Columns correspond to different lines of sight, from the left to the right:
along the major, intermediate, and minor axis of the galaxy. For clarity of the figure, in each panel we indicate
only the error bars for one of the data sets. However, as the number of stars in a sample remains roughly constant
between the lines of sight, the error bars are very similar among the panels in a given row.

Although in our previous studies of the reliability of the Schwarzschild modeling and its
applications to real data \citep{kowalczyk_2017, kowalczyk_2018, kowalczyk_2019} we approximated the density profile of
the tracer with the S\'ersic formula, we found that it does not provide a good approximation of the data for the
simulated galaxy considered here. We therefore fit the projected density profile with the King formula
\citep{king_1962}
\begin{equation}
 I(R)=I_0 \left[\frac{1}{\sqrt{1+(R/R_{\rm c})^2}}-\frac{1}{\sqrt{1+(R_{\rm t}/R_{\rm c})^2}} \right]^2,
\end{equation}
where $I_0$, $R_{\rm c}$, and $R_{\rm t}$ are the model parameters.
The profile can be analytically deprojected to obtain the 3D density
\begin{equation}
 \rho(r)=\frac{\rho_0}{z^2} \left[ \frac{1}{z}\arccos(z)-\sqrt{1-z^2} \right],
\end{equation}
where
\begin{equation}
 \rho_0=\frac{I_0}{\pi R_{\rm c} [1+(R_{\rm t}/R_{\rm c})^2]^{3/2}}
\end{equation}
and
\begin{equation}
 z=\sqrt{\frac{r^2+R_{\rm c}^2}{R_{\rm c}^2+R_{\rm t}^2}}.
\end{equation}

\section{Schwarzschild modeling}
\label{sec:model}

In this section we briefly present our modeling method and its application to the data sets derived for all stars and
the two populations of the simulated galaxy separately. In both cases our aim was to recover the profiles of the
total mass and the velocity anisotropy.

\subsection{Overview of the method}
\label{sec:method}

We follow the approach introduced in \citet{kowalczyk_2018}, namely we  model the total mass profile with the
mass-to-light ratio $\Upsilon$ varying with radius:

\begin{equation}
\label{eq:Upsilon}
\log\Upsilon(r) = \left\{
\begin{array}{ll}
\log(\Upsilon_0) & r\leq r_0\\
a(\log r - \log r_0)^c + \log(\Upsilon_0) & r> r_0\\
\end{array} \right.
\end{equation}
where $r$ is the distance from the center of the galaxy, $r_0$ is a constant, while $\Upsilon_0$, $a$, and $c$ are the
parameters of a model. We have assumed $\log r_0=0.33$ which corresponds to three softening scales for stellar
particles in the Illustris simulation.

We probed the parameter $a\in[0:1.3]$ with a step $\Delta a=0.04$ and $c\in[1.1:2.9]$ with a step $\Delta c=0.2$,
imposing the requirement on the total density profile to be monotonically decreasing with radius. For each set of
parameters and for each line of sight we generated 1200 orbits using 100 values of energy (expressed with the radius
of a circular orbit) spaced logarithmically and 12 values of the relative angular momentum spaced linearly. The outer
radius of the orbit library, that is the apocenter of the most extended orbit, was set to $r_{\rm out}=165$\,kpc in
order to cover over 0.999 of the total stellar mass based on the fitted King profile parameters.

We fit the kinematics weighted with the fraction of mass with the constrained least squares algorithm where
different values of $\Upsilon_0$ were obtained with a simple transformation of velocities given by Eq.\,12, 13, and 15
in \citet{kowalczyk_2018}. In order to smooth out the numerical artifacts, the three-dimensional $\chi^2$ spaces were
then interpolated with 12-order polynomials ($\sim a^4 c^4 \Upsilon_0^4$) that were further used to determine the
global minimums (identified as the best-fitting models) and 1, 2, 3\,$\sigma$ confidence levels which for three
parameters correspond to $\Delta \chi^2=3.53,\,8.02,\,14.2$ \citep{NR}.

\subsection{Application to mock data}
\label{sec:fit}

In the following we present the direct and inferred results of the Schwarzschild modeling of the data sets described
in Section~\ref{sec:obs}.

First, Fig.\,\ref{fig:chi} shows the distribution of the absolute values of the $\chi^2$ as a function of three
parameters of the mass-to-light ratio. In order to avoid unnecessary repetitions, we include only the plot for
the mock data obtained by observing the Illustris galaxy along its major axis as the others are qualitatively
similar. The four panels refer to fits for all stars (top left), the metal-rich population I (bottom left),
the metal-poor population II (bottom right), and the one named "populations" (top right) which is the algebraic sum of
values for both populations.

As our parametrization of the mass-to-light ratio is not intuitive we present its profiles explicitly in the first rows
of the left- and right-hand side panels of Fig.\,\ref{fig:ups} for the results obtained for all stars and the
populations, respectively. We further calculate the total density (second rows) and the total mass content (third
rows). We include the obtained orbit anisotropy within the modeled range in the bottom rows. The consecutive columns
present the results for the observations along the major, intermediate, and minor axis. Green lines indicate values
for the best-fit models whereas the colored areas of decreasing intensity correspond to 1, 2, and 3\,$\sigma$ confidence
regions obtained as extreme values allowed by the models with $\chi^2$ within a given region. In each panel the true
values from the simulation are presented with black lines while thin vertical lines mark the values of $r_0$ and the
outer range of the data sets beyond which the reliability of results drops significantly. The true mass-to-light ratio
profile was obtained by dividing the total mass by the fitted King profiles, therefore the drop at 100\,kpc is the
numerical artifact occurring at the very outskirts of the galaxy.

Whereas in  the right-hand side panels of Fig.\,\ref{fig:ups} the resulting anisotropy is obtained from the fit of all
stars and uses only the location of global minimum and confidence levels from two populations (as in the top right
panel of Fig.\,\ref{fig:chi}), in Fig.\,\ref{fig:beta_fit} we present another method of calculating the anisotropy. In
the second and third row we show the derived profiles for population I and II separately and combine them as stellar
mass weighted average in the top row. As in previous figures, three columns refer to the different lines of sight
whereas the narrow fourth one shows the behavior of the true profiles outside the modeled range which, as we noticed in
our previous studies, in a limited way influences the results. Such an impact is understandable since the stars at
larger distances from the center are still included in the line-of-sight measurements.

\subsection{Comparison of fitting results}
\label{sec:comparison}

The main strength of the two populations method comes from tracing the underlying gravitational potential at different
scales.
As can be seen in the bottom panels of Fig.\,\ref{fig:chi}, population I, which is more concentrated, is also more
sensitive to $\Upsilon_0$, but gives weaker constraints on $a$ or $c$. On the other hand, population II attempts to
reproduce the total mass content at larger distances as well, therefore showing stronger coupling between the
parameters.

The global minimums of the $\chi^2$ distributions for both approaches, that is modeling one and two populations, which
we identify as the best-fitting models, closely coincide showing that there is no internal bias in the improved method.
However, significant differences can be observed when comparing the confidence levels, mainly at $1$ and $3\,\sigma$.
Namely, we find that using two populations, the constraints we obtain on the density and anisotropy profile are much
stronger.

Additionally, the more accurate method allows us to study other effects and biases, for example the consequences of the
nonsphericity of the modeled object. Whereas for the fit of all stars the true values of the density, mass, and
anisotropy profiles are contained within $1\,\sigma$ confidence regions, the results for the populations are more or
less biased depending on the axis. They are well reproduced for the observation along the intermediate axis,
for which the effects of nonsphericity seem to cancel out, and more biased for the remaining lines of sight. We
notice a trend from under- to overestimation of the anisotropy when going from the major to the minor axis.

\section{Modeling Fornax dSph}
\label{sec:fornax}

In this section we present the application of our Schwarzschild modeling scheme to the observational data for the
Fornax dSph galaxy obtained by \citet{delpino_2015} and \citet{delpino_2017}. This study is a follow-up of the
work of \citet{kowalczyk_2019} and can be directly compared to the results presented there. Moreover, we
refer the reader to these previous publications for details on the origin of data and our procedures used for cleaning
the spectroscopic sample.

\begin{figure}[ht!]
\begin{center}
\includegraphics[width=\columnwidth]{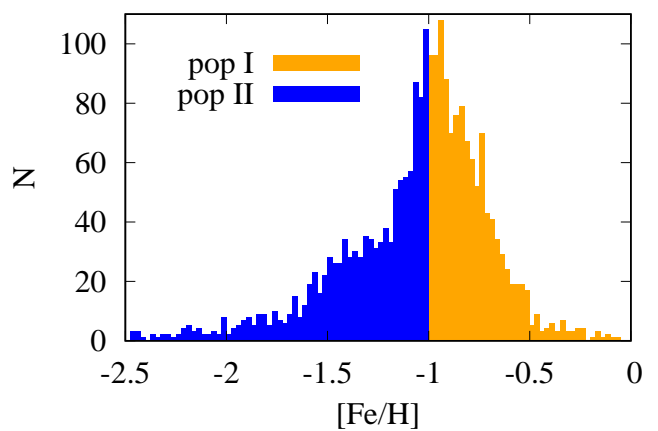}
\caption{Metallicity histogram of the final spectroscopic sample used in the modeling of two stellar populations
in the Fornax dSph. Each bin is color-coded according to the population it has been assigned to, orange or blue for
population I and II, respectively.}
\label{fig:metal_fornax}
\end{center}
\end{figure}

Similarly to the approach introduced in Section~\ref{sec:pops}, we divided all available stars into two equal-size
populations based on their metallicity and then cross-correlated the samples with the data used in
\citet{kowalczyk_2019}. The metallicity histogram of the final spectroscopic sample is shown in
Fig.\,\ref{fig:metal_fornax}. Additionally, we color-coded each bin with the population it has been assigned to, namely
orange or blue for population I or II. Interestingly, the case of Fornax is similar to our simulated galaxy as the
split at [Fe/H]$=-1$ also captures an important feature of the object's star formation history, separating stars into
subsamples older and younger than 6 Gyr, as shown in Fig. 12 of \citet{delpino_2015} and Fig. 8 of
\citet{delpino_2017}. The numbers of stars contained in the samples of all stars, population I, and population II are
given in Table\,\ref{tab:fornax}, where the indices "phot" and "spec" refer to the photometric and kinematic samples.
The sum of stars in the populations is lower than in the sample of all stars since only stars with reliable
measurements of metallicity could be included.

\begin{table}
\caption{Properties of the data samples for the Fornax dSph.}
\label{tab:fornax}
\centering
\begin{tabular}{l c c c}
\hline\hline
Property & ALL & POP I & POP II  \\ \hline
Number of stars ($N_{\rm phot}$) & 65\,797 & 14\,882 & 49\,205 \\
Number of stars ($N_{\rm spec}$) & 3286 & 1136 & 1151 \\
Stars within 1.8\,kpc & 3268 & 1134 & 1130 \\
\hline
Fitted normalization ($N_0$) [$\times 10^4$] & 6.95 & 1.81 & 5.45 \\
S\'ersic radius ($R_{\rm S}$) [kpc] & 0.454 & 0.429 & 0.420 \\
S\'ersic parameter ($m$) & 0.808 & 0.807 & 0.898 \\
\hline
\end{tabular}
\end{table}

\begin{figure}[ht!]
\begin{center}
\includegraphics[width=\columnwidth]{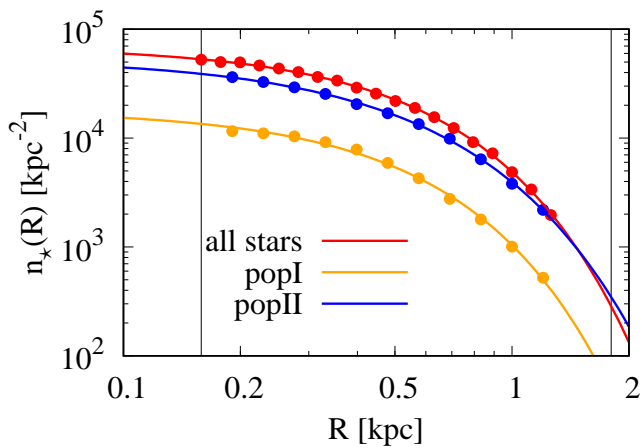}
\caption{Surface number density profiles of the photometric data samples for the Fornax dSph: all available stars (in
red), the metal-rich population I (in orange), and the metal-poor population II (in blue). Thin vertical lines indicate
$r_0$ (see text) and the outer boundary of the spectroscopic data.}
\label{fig:niu_fornax}
\end{center}
\end{figure}

\begin{figure}[ht!]
\begin{center}
\includegraphics[trim=35 0 0 100, clip, width=\columnwidth]{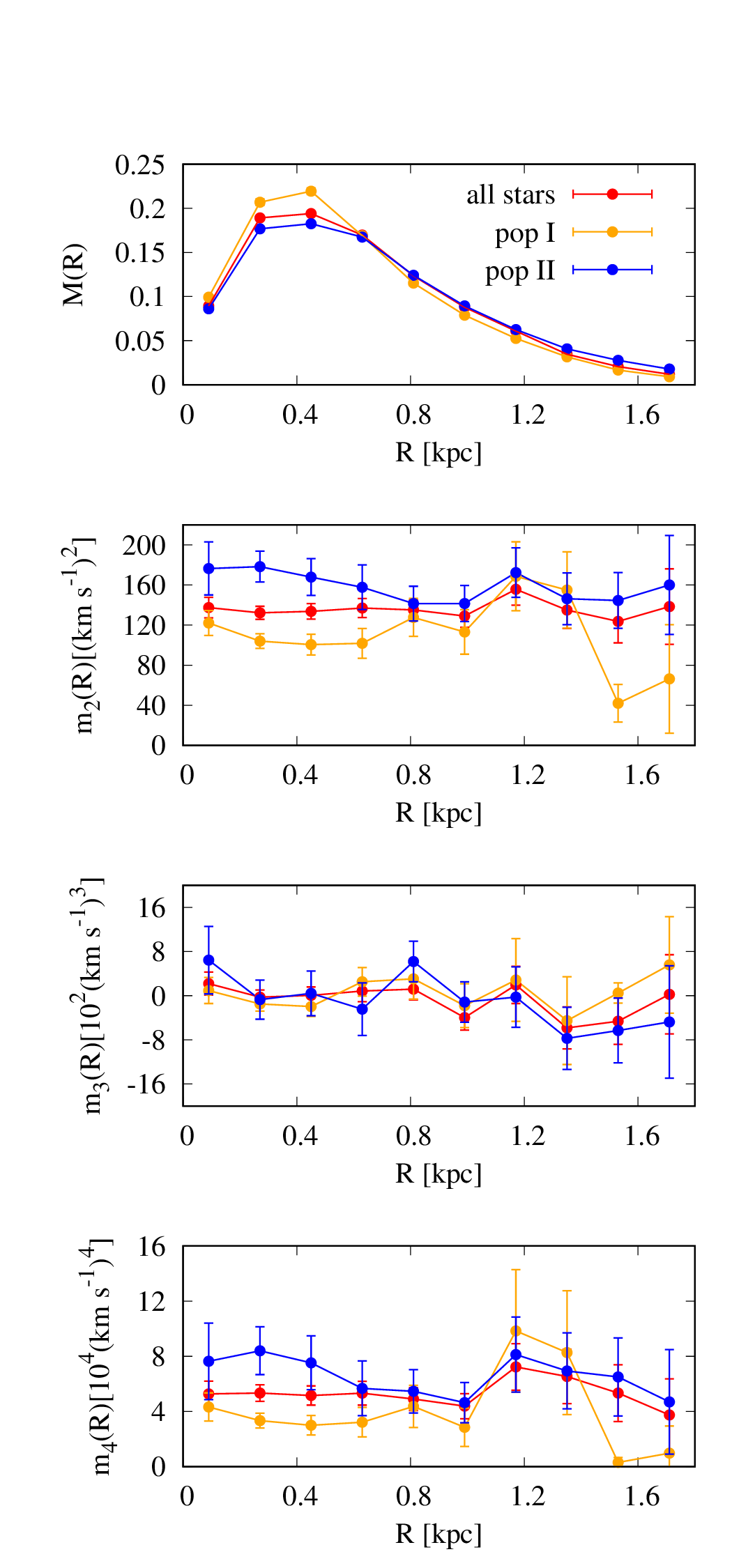}
\caption{Observables of the Fornax dSph used in our Schwarzschild modeling scheme. \textit{In rows:} the fraction of
the total number of stars, the 2nd, 3rd, and 4th velocity moment. In red we present the values obtained for all stars
whereas in orange and blue those for populations I and II, respectively.}
\label{fig:obs_fornax}
\end{center}
\end{figure}

As we have shown in our earlier work, the light profile of the Fornax dSph can be well reproduced with the
three-parameter S\'ersic formula \citep{sersic_1968}. The profiles of number density for all stars and both populations
together with the best-fitting S\'ersic profiles are presented in Fig.\,\ref{fig:niu_fornax}. The colors follow the
convention introduced in previous sections. Thin vertical lines indicate the innermost data point for the light profile
for all stars and the outer boundary of the kinematic sample. The former, set at $\log r=-0.16$, is also used as the
minimum of the mass-to-light ratio profile ($r_0$ in Eq.\,\ref{eq:Upsilon}). The fitted parameters of the profiles,
that is the normalization $N_0$, the S\'ersic radius $R_{\rm S}$, and the S\'ersic parameter $m$, are included in the
second part of Table~\ref{tab:fornax}.

Figure\,\ref{fig:obs_fornax} presents the profiles of the observables used in the Schwarzschild modeling: the fraction
of stars and the 2nd, 3rd, and 4th velocity moments (top to bottom) for the three data samples: all stars, population
I, and population II (in red, orange, and blue, respectively). The error bars indicate 1\,$\sigma$ sampling errors.

\begin{figure*}[ht!]
\centering
\includegraphics[trim=0 1100 0 0, clip, width=0.9\textwidth]{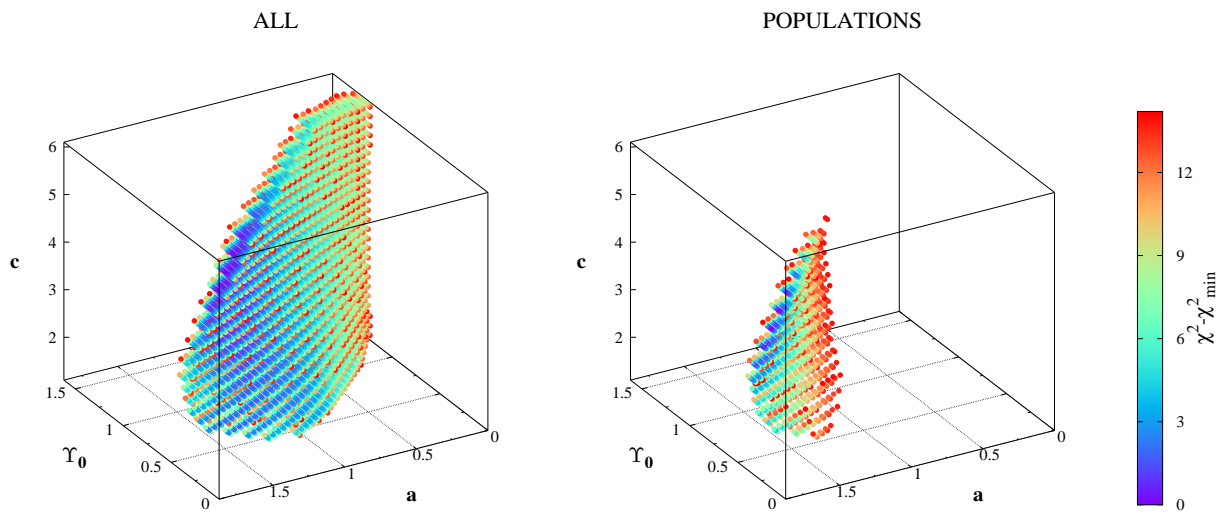}
\caption{Values of $\chi^2$ relative to the fitted minimum within the range of $3\,\sigma$ confidence level for all
stars (left panel) and for the populations (right panel) for the Fornax dSph.}
\label{fig:chi_fornax}
\end{figure*}

\begin{figure}[ht!]
\begin{center}
\includegraphics[trim=0 0 0 0, clip, width=\columnwidth]{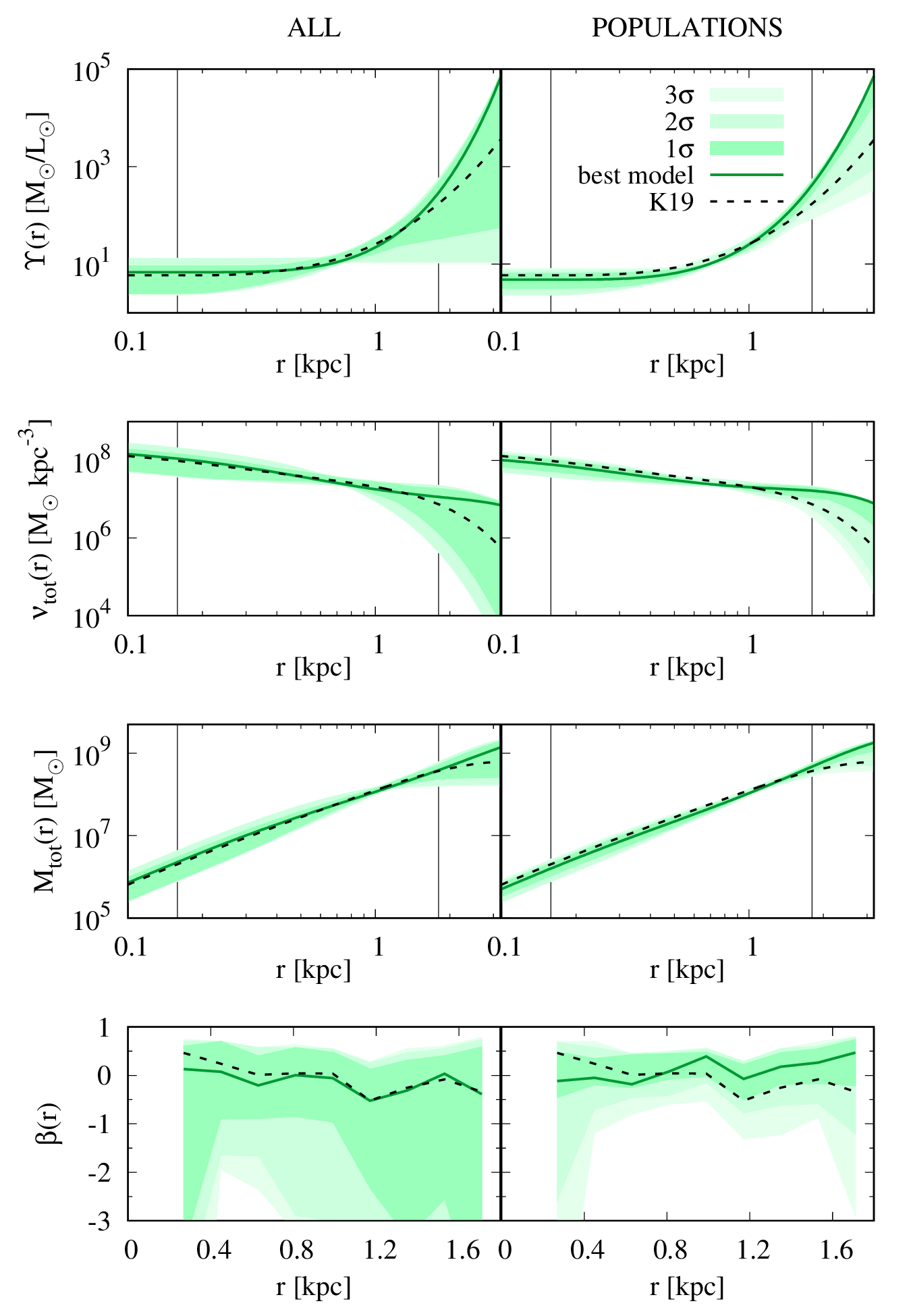}
\caption{Results of Schwarzschild modeling of the Fornax dSph. \textit{In rows:} derived mass-to-light ratio, total
density, total mass, and anisotropy parameter. \textit{In columns:} results for all stars and the populations,
respectively. Green lines indicate the values for the best-fit models whereas the colored areas of decreasing intensity
show the 1, 2, and 3\,$\sigma$ confidence regions. The best-fitting values obtained by
\citet{kowalczyk_2019} are shown with black dashed lines.}
\label{fig:ups_fornax}
\end{center}
\end{figure}

\begin{figure}[ht!]
\begin{center}
\includegraphics[width=0.8\columnwidth]{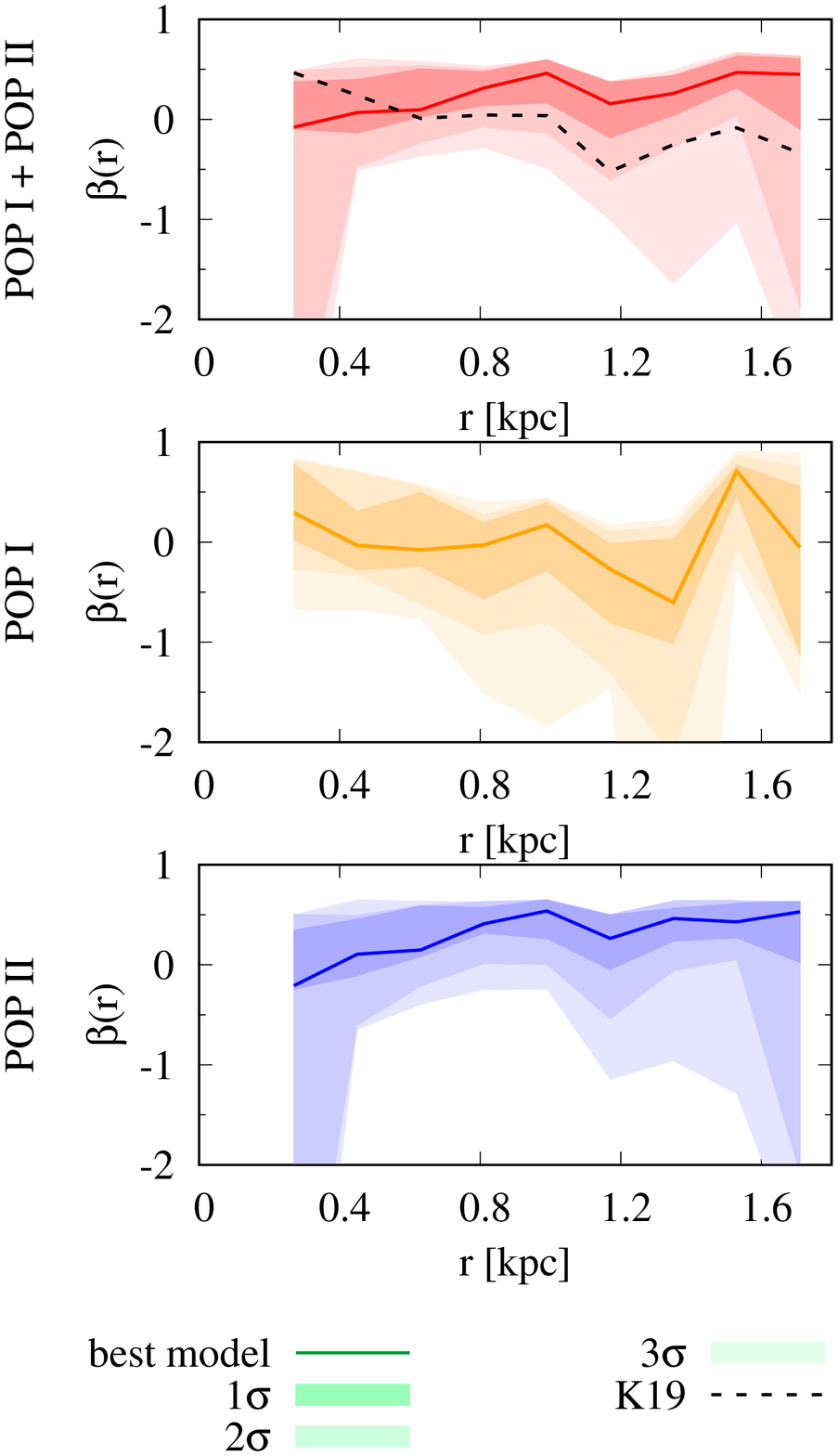}
\caption{Profiles of the anisotropy parameter obtained with the Schwarzschild modeling of two stellar
populations for the Fornax dSph. \textit{In rows:} results for all stars (calculated as the superposition of two
populations), population I, and population II. Color lines indicate values for the best-fit models whereas the colored
areas of decreasing intensity show the 1, 2, and 3\,$\sigma$ confidence regions. The dashed black line shows the result
from \citet{kowalczyk_2019} for comparison.}
\label{fig:beta_fornax}
\end{center}
\end{figure}

The parameter space for $\Upsilon (r)$ has been probed as follows: $a\in[0:1.85]$ with a step $\Delta a=0.05$ and
$c\in[1.2:6]$ with a step $\Delta c=0.2$. We point out that in \citet{kowalczyk_2019} the parameter $c$ was fixed at
$c=3$ and now we fit it as a free parameter. As for the mock data in Section~\ref{sec:fit}, different values of
$\Upsilon_0$ were obtained with the transformation of velocity moments within the $\chi^2$ fitting routine.
The values of $\Delta \chi^2$ for all stars and the populations are shown in the two panels of
Fig.\,\ref{fig:chi_fornax} (left and right-hand side, respectively). Due to the dense coverage of the grid, we decided
to include only the values within 3\,$\sigma$ from the fitted minimums (see Section~\ref{sec:method}).

The profiles of the mass-to-light ratio, total density, total mass, and velocity anisotropy resulting from the $\chi^2$
distributions are presented in the consecutive rows of Fig.\,\ref{fig:ups_fornax}. The anisotropy profile for the
populations is based on the fit of all stars but using the confidence levels on $\Upsilon$ from the fit of two
populations. Green lines indicate the values for the best-fitting models whereas the colored areas of decreasing
intensity show the 1, 2, and 3\,$\sigma$ confidence regions. Additionally, with black dashed lines
we include the results from \citet{kowalczyk_2019} for comparison.

As a result of freeing the steepness of the mass-to-light ratio profile (parameter $c$) with respect to the previous
study \citep{kowalczyk_2019}, we obtained higher estimates of the enclosed total mass at larger radii. In particular,
for the mass enclosed within 1.8 kpc we get $M_{\rm all}(<1.8 \,$kpc$)=3.87_{-1.56}^{+1.48} \times 10^8$ M$_\odot$ from
the fit for all stars and $M_{\rm pops}(<1.8 \,$ kpc$)=4.71_{-1.13}^{+0.87} \times 10^8$ M$_\odot$ from the fit of
populations, while previously we had $M_{\rm old}(<1.8 \,$ kpc$)=3.7_{-1.3}^{+1.4} \times 10^8$ M$_\odot$.

Interestingly, despite the significant shift of the position of $\chi^2_{\rm min}$ (to $c=4.2$ for all stars and 3.6
for populations), the obtained profile of the anisotropy parameter remains decreasing or flat for all stars but changes
to increasing from 0 to 0.5 for the populations. Nevertheless, even in the latter case the previous result agrees with
the new finding within 1\,$\sigma$.

The detailed analysis of the anisotropy is shown in Fig.\,\ref{fig:beta_fornax} where the middle and bottom panels
present the profiles obtained for each population separately. We notice that the profile for population I is decreasing
or has a local minimum whereas for population II is increasing (from $-0.25$ to 0.5 for the best-fitting model). Since
population I is more concentrated, the last bins contain very few stars, which limits their credibility. The top panel
of Fig.\,\ref{fig:beta_fornax} presents the anisotropy of all stars calculated as a weighted superposition of two
populations. With such approach we still obtain the increasing profile (from 0 to 0.5) but the previous result agrees
with it only within 2\,$\sigma$.

Since Fornax dSph is significantly elongated with the projected ellipticity of $\epsilon=0.30\pm 0.01$
\citep{irwin_1995}, we anticipate some bias in the obtained results caused by the spherically symmetric modeling.
\citet{kowalczyk_2018} studied such bias in an axisymmetric simulated object qualitatively similar to Fornax and
identified differences in the systematic errors depending on whether the galaxy was observed along its major or
minor axis. Assuming that Fornax is observed along the line of sight in between these extremes, we expect the total
mass profile to be slightly overestimated and the anisotropy to be underestimated, further strengthening the likelihood
of the real anisotropy to be radial and its profile to be growing with radius with respect to the results of
\citet{kowalczyk_2019}.

Both constant (like for our population I) and growing (population II) anisotropy profiles can arise from biased
modeling of the real growing profile by observing an object along the minor and major axis, respectively. However, for
the bias to occur in two populations presented here, their inner orientations would need to be opposite. Since such
morphological features are not supported by the photometric studies of Fornax \citep{delpino_2015, wang_2019} which
rather find a good spatial alignment between the stellar populations, we conclude that the anisotropy profiles of the
two populations modeled in this work are indeed significantly distinct.

Finally, it is worth noticing that the so-called mass-follows-light model, that is the one following from the
assumption that the total density traces the stellar distribution, is no longer supported by the fit of the populations.
With our parametrization, the mass-follows-light model corresponds to $a=0$ and whereas it is enclosed within
3\,$\sigma$ for the fit of all stars, as was the case in \citet{kowalczyk_2019}, the allowed values for the
improved method are much larger, as demonstrated by the right panel of Fig.~\ref{fig:chi_fornax}.

\section{Summary and discussion}
\label{sec:summary}

Building on the previously created implementation of the Schwarzschild orbit superposition method focused on
modeling dSph galaxies of the Local Group \citep{kowalczyk_2017, kowalczyk_2018, kowalczyk_2019}, we
improved our tool by introducing multiple stellar populations. Such an improvement is desirable and justified since
many of the dwarfs show signs of multiple star formation bursts or extended star formation episodes. As the
different populations trace the common underlying gravitational potential, one may expect a significant improvement in
the estimates of not only the total mass content but also the orbit anisotropy since this robust modeling technique
reproduces the anisotropy as a by-product of the modeling rather than taking it as an assumption.

We have tested our hypothesis by modeling mock data generated from a galaxy formed in the Illustris simulation. Due to
the limitations of the resolution, we chose a galaxy of mass a few orders of magnitude larger
than the estimated masses of classical dwarfs. Still, the galaxy possessed appropriate qualitative characteristics,
such as the lack of gas and an almost spherical shape, that made it a good test bed for modeling techniques applicable
to dSph galaxies. We applied our approach to all data and to two stellar populations separately, comparing the
accuracy of the obtained results. Although the addition of the second tracer seemingly increases the number of
constraints twice, the increment is somewhat compromised by the sampling errors since the number of stars in each
sample is then reduced. Still, we found strong improvements in the accuracy of the method when using two
populations. The results of the modeling show that the density and velocity anisotropy profiles are more strongly
constrained, most importantly at the 3\,$\sigma$ level, that is the range of allowed values is much narrower.

Similarly to the conclusions of \citet{kowalczyk_2018} who explored the effects of nonsphericity using large
and small data samples, the comparison of results presented in the left- and right-hand side panels of
Fig.\,\ref{fig:ups} suggests that the improved method using two stellar populations gives more precise but less accurate
outcome. However, in both studies the apparent deterioration of the reliability is a consequence of modeling of a
nonspherical object. In both cases, a simpler approach (much smaller data samples or using one stellar population)
resulted in larger final uncertainties, usually containing the true values within $1\,\sigma$ confidence
region. On the other hand, the improved methods exhibit substantially reduced uncertainties, highlighting the underlying
bias.

Our method parametrizes the total mass content with the mass-to-light ratio varying with radius as a power-law in the
log-log scale. We made two main changes with respect to our previous work: we added a third parameter $c$ controlling
the steepness of the mass-to-light ratio profile (previously fixed at the value of 3) and allowed for different stellar
density profiles (previously only S\'ersic, now also King). These changes are of course coupled since different density
profiles require different exponents to reproduce the same mass profile. It is visible also in our results since
the King profile applied in the simulated galaxy gave us values of $c$ lower than 3. Nevertheless, we decided to use
different density profiles to make our method more general and applicable to objects, such as our Illustris galaxy, for
which the S\'ersic formula does not provide a good approximation of the density distribution.

Finally, we applied the improved method to the data for the Fornax dSph galaxy. Due to the addition of another free
parameter in our functional form for the mass-to-light ratio, our results for modeling all stars are slightly
different from the ones obtained in \citet{kowalczyk_2019}. However, in terms of the total density and mass
distribution the estimates obtained here agree very well with those earlier results in the range covered by the data.
Therefore, the detailed comparison with other estimates from the literature presented in \citet{kowalczyk_2019} is
still valid and we do not repeat it here.

A more significant difference with respect to these previous estimates is seen in the results of modeling two
populations in Fornax. In this case we find the anisotropy to be slightly increasing rather than decreasing with radius
and, most importantly, the confidence regions for this parameter, as well as for the density, are much narrower. We were
thus able to obtain tighter constraints on the properties of Fornax, which means that the improved
method is successful. For the first time, we were also able to deduce the velocity anisotropy profiles for each of the
populations separately. We found that the more concentrated, metal-rich population I has a decreasing anisotropy
profile while the more extended, metal-poor population II has the anisotropy increasing with radius.
This finding may partially explain the large spread of the anisotropy values obtained in the literature
and summarized in Table 2 and 3 of \citet{kowalczyk_2019}, which were often based on modeling subsamples of our
spectroscopic data set.

For both studied objects we split the stars into two populations by dividing them in half based on their
metallicity, $Z$ (in solar units), for the Illustris galaxy and [Fe/H] for Fornax. Such a method is approximate but
justified. Both galaxies have complex star formation history with multiple star formation bursts, as
demonstrated by Fig.\,\ref{fig:sfr} in this work and Fig. 7 in \citet{delpino_2013}, producing multiple
stellar populations which cannot be easily tracked as the metallicity is a good but not perfect proxy for the stellar
age. Moreover, the metallicity histograms for both objects are approximately unimodal not allowing for a convenient
separation. More refined methods of division have been suggested in the literature, for example in the form of the
likelihood function based on the position, velocity, and metallicity index \citep{walker_2011}. However, the likelihood
function requires many assumptions which introduce additional uncertainties into the treatment of the data. On the other
hand, our approach ensures the maximization of each sample (and therefore minimization of sampling errors) while
capturing the important features of the star formation history.

Further improvements to the Schwarzschild modeling method are certainly possible. One way to proceed would be to
include the modeling of the proper motions of the stars. For now, measurements of transverse velocities are available
only for the brightest stars in dSph galaxies, but even small samples of this type could provide further constraints
on the models, as demonstrated by \citet{strigari_2007} and \citet{massari_2020}.

\begin{acknowledgements}
We are grateful to Andr\'es del Pino for providing the data for the Fornax dSph and to the Illustris
team for making their simulations publicly available. Useful comments from the anonymous referee are kindly
appreciated. This research was supported by the Polish National Science Center under grant 2018/28/C/ST9/00529.
\end{acknowledgements}

\end{document}